\documentclass[twocolumn,showpacs,prb]{revtex4}
\usepackage{bm}
\usepackage{epsfig}
\usepackage{graphicx}

\newcommand{\nix}[1]{}
\begin{document}

\title{
Rashba and Dresselhaus Spin-Splittings
in Semiconductor Quantum Wells Measured by Spin Photocurrents}
\author{S.~Giglberger$^1$, L.E.~Golub$^2$,  V.V.~Bel'kov$^2$, S.N.~Danilov$^{1}$, D.~Schuh$^1$, Ch.~Gerl$^{1}$,
F.~Rohlfing$^{1}$,  J.~Stahl$^{1}$, W.~Wegscheider$^1$, D.~Weiss$^1$,
W.~Prettl$^1$, and S.D.~Ganichev$^{1}$
}
\affiliation{$^1$Fakult\"{a}t Physik, University of Regensburg,
93040, Regensburg, Germany}
\affiliation{$^2$A.F.~Ioffe Physico-Technical Institute, Russian
Academy of Sciences, 194021 St.~Petersburg, Russia}

\date{\today}

\begin{abstract}

The spin-galvanic effect and the circular photogalvanic effect
induced by terahertz radiation are applied to determine the
relative strengths of Rashba and Dresselhaus band spin-splitting
in (001)-grown GaAs and InAs based two dimensional electron
systems. We observed that shifting the $\delta$-doping plane  from
one side of the quantum well to the other results in a change of
sign of the photocurrent caused by  Rashba spin-splitting while
the sign of the Dresselhaus term induced photocurrent remains. The
measurements give the necessary feedback for technologists looking
for structures with equal Rashba and Dresselhaus spin-splittings
or perfectly symmetric structures with zero Rashba constant.

\end{abstract}

\maketitle

\section{Introduction}

In low dimensional structures based on  III-V compound
semiconductors the spin degeneracy of the energy bands  is removed.
This lifting of spin degeneracy is caused by spin-orbit
interaction and results  in terms linear in electron wavevector
${\bm k}$ in the effective Hamiltonian.  The spin-splitting is
crucial for the field of spintronics, indeed it allows  the
electric field control of spin polarization, determines the spin
relaxation rate, and can be utilized for all-electric spin
injection.~\cite{spintronicbook02} The microscopic origin of 
terms linear in electron wavevector in low dimensional systems is
structure inversion asymmetry (SIA) and bulk inversion asymmetry
(BIA) which lead to Rashba and Dresselhaus spin-orbit terms in the
Hamiltonian, respectively.~\cite{Bychkov84p78,Dyakonov86p110}
These terms can interfere resulting in an anisotropy of spin
splitting and can even cancel each other if Rashba and Dresselhaus
terms have equal strength resulting in a vanishing spin splitting
in certain {\boldmath$k$}-space directions.~\cite{review2003spin}
This cancellation leads to new macroscopic effects such as disappearance
of anti-localization,~\cite{Knap1996p3912} the absence of spin
relaxation in specific crystallographic
directions,~\cite{Averkiev1999p15582,Averkiev02pR271}  the lack of
Shubnikov-de Haas beating,~\cite{Tarasenko02p552} and can be
employed for a non-ballistic spin-field effect
transistor.~\cite{Schliemann03p146801} Thus, the knowledge of the
relative strength of  Rashba and Dresselhaus terms is important
for investigations of spin dependent phenomena in low dimensional
structures. Recently we demonstrated~\cite{PRL04} that the
spin-galvanic effect
(SGE)~\cite{Nature02,GanichevPrettl,Ivchenkobook2}
can be used as
a tool to measure
the ratio of Rashba and  Dresselhaus terms (R/D-ratio)  in a
very direct way which does not rely on theoretically obtained
quantities. Investigation of the angular dependence of the
spin-galvanic current in InAs quantum wells, excited by terahertz
radiation, allowed us to deduce the R/D-ratio directly from
experiment. In the present paper we apply this method to various (001)-oriented $n$-type
InAs- and GaAs-based heterostructures
and extended  this novel technique by employing the circular photogalvanic effect
(CPGE).~\cite{review2003spin,GanichevPrettl,Ivchenkobook2,PRL01,Sasaki03,Bieler05,Yang06}

\section{Spin photocurrents as a method}

A direct way to explore the BIA and SIA terms in the Hamiltonian,
which requires no knowledge of microscopic details, is based on the
phenomenological equivalence of different mechanisms linearly
coupling  a polar vector such as wavevector or current with an
axial vector like spin of electrons  or angular momentum of
photons. Indeed, such phenomena are described by  second rank
pseudo-tensors whose irreducible components differ by a scalar
factor only. Thus, the anisotropy in space
is the same for all such phenomena. The strength of the spin
splitting in various crystallographic directions is described by
Rashba and Dresselhaus terms in the Hamiltonian
$H_{SO}=\sum\beta_{lm}\sigma_lk_m$, where $\beta_{lm}$ is a second
rank pseudo-tensor and $\sigma_l$ is the Pauli spin matrix. 
Non-zero components of second rank
pseudo-tensors may exist in gyrotropic point groups only, where,
by definition,
axial and polar vector components transform
equivalently by all symmetry operations. Thus the linear coupling
between these vectors becomes possible.
There are two experimentally accessible effects which are also
described by such second rank pseudo-tensors: the spin-galvanic
effect,\cite{Nature02} $j_\alpha = \sum Q_{\alpha \beta} S_\beta$, and the
circular photogalvanic effect,\cite{PRL01} $j_\alpha=\sum \gamma_{\alpha \beta}
P_{\rm circ} \hat{e}_\beta $. Here $\bm S$ is the average spin,
$P_{\rm circ}$ is a circular polarization degree, and $\hat{\bm
e}$ is the projection of the unit vector pointing in the direction
of light propagation  onto the plane of the sample. In analogy to
the band spin-splitting and based on the equivalence of the
invariant irreducible components of the pseudotensors ${\bm \beta,
Q}$  and ${\bm  \gamma}$, these currents can be decomposed into
Rashba and Dresselhaus contributions which can be measured
separately. Taking the ratio between these contributions cancels
the scalar factor which contains all microscopic details.\cite{PRL04}
Therefore the ratio is constant for all mechanisms and
photocurrents and can be used to determine the anisotropy of the spin
splitting which is otherwise experimentally not easily accessible.

\subsection{Band structure}

First we briefly summarize the consequences of Rashba and
Dresselhaus terms on the electron dispersion and on the spin
orientation of the 2DEG's electronic eigenstates. We consider QWs
of zinc-blende structure grown in [001] direction, belonging to
C$_{2v}$ point-group symmetry. In this point group the
pseudo-tensor $\beta_{lm}$ has two non-vanishing invariant
irreducible components, $\alpha=\beta_{xy}=-\beta_{yx}$ and $\beta
= \beta_{xx} = -\beta_{yy}$ which yield  Rashba and  Dresselhaus
terms, respectively. The coordinate system used here is along
the cubic axes of the crystal, $x \parallel [1 0 0]$, $y
\parallel [0 1 0]$.
The corresponding spin-orbit part ${H}_{SO}$ of the ground electron subband Hamiltonian,
$${H} = \hbar^2 k^2 / 2m^* + {H}_{SO},$$
($m^*$ is the effective electron mass) contains a Rashba term as well as a Dresselhaus term according to
\begin{equation}\label{H}
H_{SO} =  \alpha (\sigma_{x} k_{y} - \sigma_{y} k_{x}) + \beta
(\sigma_{x} k_{x} - \sigma_{y} k_{y}) \,\, .
\end{equation}

The energy spectrum of such systems consists of two branches with
the following anisotropic dispersions~\cite{Silva92p1921}
\begin{equation}\label{spectrum}
    \varepsilon_\pm({\bm k}) = {\hbar^2 k^2 \over 2 m^*} \pm
    k\sqrt{\alpha^2 + \beta^2 + 2 \alpha \beta \sin{2 \vartheta_{\bm k}}} \,\, ,
\end{equation}
where $\vartheta_{\bm k}$ is the angle between $\bm k$ and the $x$ axis.

\begin{figure}[h!]
	\centerline{\epsfxsize 86mm \epsfbox{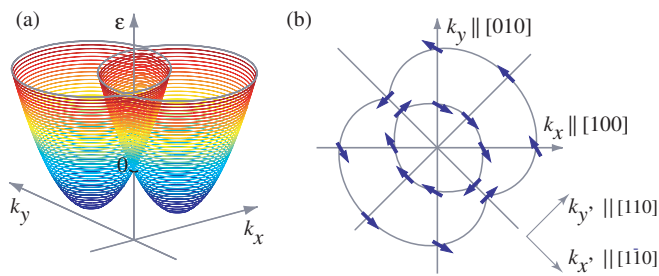}}
	\caption{Schematic 2D  band structure with $\bm{k}$-linear terms
	for C$_{2v}$ symmetry for non-equal strength of SIA and BIA. (a)
	The energy $\varepsilon$ as a function of $k_x$ and $k_y$, (b) the
	distribution of spin orientations at the 2D Fermi energy.}
	\label{fig1}
\end{figure}

To illustrate the resulting energy dispersion we plot in
Fig.~\ref{fig1} the eigenvalues of ${H}$,
$\varepsilon$({\boldmath$k$}) (left panel), and contours of
constant energy in the $k_{x}$,$k_{y}$ plane together with the
spin orientation of eigenstates at selected points in
{\boldmath$k$}-space (right panel).
Here we assumed that
$\alpha > \beta >0$, and $\alpha/\beta = 1.15$.
Rashba and Dresselhaus terms result in different patterns of  spin
eigenstates in {\boldmath$k$}-space. The distribution of  spin
directions in {\boldmath$k$}-space can be visualized by writing
the spin-orbit interaction term in the form $H_{SO} = \hbar
\mbox{\boldmath$\sigma$} \cdot \mbox{$\bm {\Omega}$}_{\rm
eff}(\bm{k})$, where $\mbox{$\bm {\Omega}$}_{\rm eff}(\bm{k})$ is
an effective Larmor precession frequency providing the
relevant quantization axis. In the presence of both Rashba and
Dresselhaus spin-orbit couplings, characteristic for C$_{2v}$
symmetry, the $[1\bar {1}0]$ and the [110] axes become
non-equivalent (see Fig.~\ref{fig1}). The maximum and minimal spin
splittings take place along these axes and are equal to
$2(\alpha+\beta)k$ and $2(\alpha-\beta)k$, respectively.

\subsubsection*{The role of ${\bm k}$-cubic spin-orbit splitting}

So far we discussed $k$-linear terms in the Hamiltonian only. In fact, in
zinc-blende structure based (001)-grown quantum wells also terms cubic in
$k$ are present which stem from the Dresselhaus term in the host bulk material.
The corresponding effective precession frequency ${\bm \Omega}_{\rm eff} \sim k^3$ in
quantum wells can be conveniently decomposed as ${\bm \Omega}_{\rm eff} =
{\bm \Omega}_1 + {\bm \Omega}_3$, where ${\bm \Omega}_1$ varies with the angle
$\vartheta_{\bm k}$ as combinations of $\cos{\vartheta_{\bm k}}$,
$\sin{\vartheta_{\bm k}}$ and ${\bm \Omega}_3$ as
$ \cos{3 \vartheta_{\bm k}}$,
$\sin{3 \vartheta_{\bm k}}$ terms (see for instance Ref.~\onlinecite{Knap1996p3912,book}).

Both SGE and CPGE are   caused by  the term ${\bm \Omega}_1$ only.
 Hence, the
Dresselhaus term in the Hamiltonian, which yields the
photocurrent, is given by the sum of the term linear in $\bm k$
discussed above and the terms cubic in $\bm k$ given by $\bm
\Omega_1$. The last terms only  renormalize the Dresselhaus
parameter  $\beta$ which should be replaced by $\beta - \gamma
k^2/4$. Here $\gamma$ is the bulk spin-orbit constant.
The cubic terms
given by ${\bm \Omega}_3$ do not result in spin photocurrents,
however, they modify the spin splitting and may affect spin
relaxation and the anisotropy of spin-flip Raman
scattering.~\cite{book,Jusserand95p4707,Miller03p076807,Lusakowski2003,Flatte}

\subsection{Spin-galvanic effect}

The spin-galvanic effect  consists of the generation of an electric current
due to a non-equilibrium  spin polarization and is caused by asymmetric  spin relaxation.
\cite{review2003spin,Nature02,GanichevPrettl,Ivchenkobook2}
The spin-galvanic effect  generally  needs no optical excitation but may also occur
due to optical spin orientation yielding a spin photocurrent.
The SGE current ${\bm j}_{SGE}$ and the average spin
 are related by a second rank
pseudo-tensor with components proportional to the parameters of spin-orbit splitting
as follows
\begin{equation}
\label{SGE} {\bm j}_{SGE}  = A \left( {{\begin{array}{*{20}c}
 \beta \hfill & -\alpha \hfill \\
  \alpha  \hfill & - \beta \hfill \\
\end{array} }} \right){\bm S} \,\, ,
\end{equation}
where ${\bm S}$ is the average spin in the plane of the
heterostructure and  $A$ is a constant determined by the kinetics
of the SGE, namely by the characteristics of momentum and spin
relaxation processes.

\begin{figure}
	\centerline{\epsfxsize 86mm \epsfbox{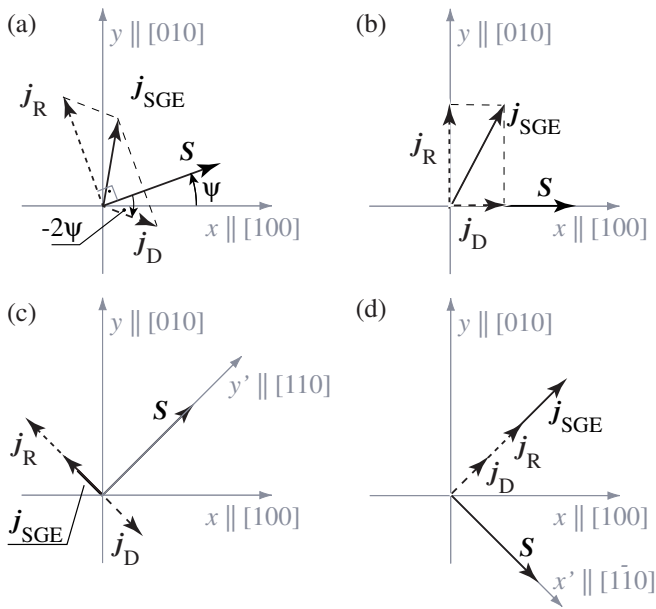}}
\caption{ Spin-galvanic currents in an (001)-grown  QW for
in-plane average spin direction given by arbitrary angle $\Psi$
(a), SGE-I geometry (b), and SGE-II geometry (c) and (d).  It is assumed that $\alpha
> \beta > 0$.}	\label{fig2}
\end{figure}

Eq.~(\ref{SGE}) allows one to determine  the R/D-ratio. This is
sketched in Fig.~\ref{fig2} (a) showing the average spin ${\bm S}$
and the spin-galvanic current ${\bm j}_{SGE}$ which is decomposed
into ${\bm j}_R$ and ${\bm j}_D$ proportional to the Rashba
constant $\alpha$ and the Dresselhaus  constant $\beta$,
respectively.
By symmetry arguments the current  ${\bm j}_R$ is always
perpendicular to the average spin ${\bm S}$ while the current
${\bm j}_D$ encloses an angle $-2\Psi$ with ${\bm S}$, where
$\Psi$ is the angle between ${\bm S}$ and the $[100]$-axis.  The
absolute value of the total current $j_{SGE}$ is given by the
expression
\begin{equation}
\label{j_total}
j_{SGE} = \sqrt{j_R^2 + j_D^2 - 2 j_R j_D \sin{2\Psi}} \,\, ,
\end{equation}
which has the same algebraic form as the spin-orbit term in the
band structure, see Eq.~(\ref{spectrum}). Hence, by mapping the
magnitude of the photocurrent in the plane of the QW the R/D-ratio
$\alpha/\beta$ can be directly extracted from experiment. Two
experimental geometries, SGE-I and SGE-II,
are particularly suited to obtain $\alpha/\beta$.

\subsubsection*{Geometry SGE-I}

In the first case ${\bm S}$ is oriented  along the $x$ axis parallel
to the $[100]$-direction ($\Psi = 0$). Then it follows from Eq.~(\ref{j_total})
that ${\bm j}_D$ and ${\bm j}_R$ are directed along and perpendicular
to ${\bm S}$, respectively. This situation is sketched in
Fig.~\ref{fig2} (b). The ratio of the currents measured along
the $x$ and the $y$ axes gives
\begin{equation}
\label{RDgeomI}
\frac{\alpha}{\beta} = \frac{j_y(\bm S \parallel x)}{j_x(\bm S \parallel x)}\,\,.
\end{equation}
Besides the ratio of two spin splitting contributions, this
geometry unambiguously shows whether the Rashba or  Dresselhaus
contribution is dominating. Furthermore, Eq.~(\ref{RDgeomI})
allows one the experimental determination of not only the ratio but also
the relative sign of the Rashba and Dresselhaus constants.

\subsubsection*{Geometry SGE-II}

Figures~\ref{fig2} (c) and~\ref{fig2} (d) illustrate the second geometry.
Here  two measurements need to be carried
out, one with  ${\bm S}$ along the $y^\prime$-axis parallel to
the [110]-direction ($\Psi=  \pi/4$), and the second with ${\bm S}$
along the $x^\prime$-axis parallel to the [1$\bar{1}$0]-direction
($\Psi= -\pi/4$). In both cases the total current flows
perpendicularly to the spin. For the spin oriented along $y^\prime$
the Rashba and Dresselhaus current contributions are oppositely
alligned and the total current  has the magnitude
$j_{x^\prime}({\bm S} \parallel y^\prime)  = j_R - j_D$. For the
spin along $x^\prime$, however, the current contributions are
collinear and the total current is given by $j_{y^\prime}({\bm S}
\parallel x^\prime) =j_R+j_D$. The relative strength of these
currents
\begin{equation}
\label{RDgeomII} r = \left|  \frac{j_{x^\prime }({\bm S} \parallel
y^\prime)}{j_{y^\prime }({\bm S} \parallel x^\prime)}  \right| =
\left| \frac{\alpha - \beta}{\alpha + \beta} \right|
\end{equation}
allows us a quantitative  determination of the R/D-ratio via
\begin{equation}
\label{RDgeomII_a} \frac{\alpha}{\beta} = \frac{1+r}{1-r} \,\, .
\end{equation}
Thus the geometry SGE-II  gives information about the relative
strength of the spin splitting parameters, however, in contrast to
the geometry SGE-I, it does not distinguish between the Rashba and
Dresselhaus terms. This fact makes such a procedure not as useful
as the geometry SGE-I. Here we use the geometry SGE-II for
demonstration of self-consistency of the method only. However, as
we show below, this geometry is suitable  for controllable growing
of structures with equal Rashba and Dresselhaus constants.



\subsection{Circular photogalvanic effect}

%

\begin{figure}
	\centerline{\epsfxsize 86mm \epsfbox{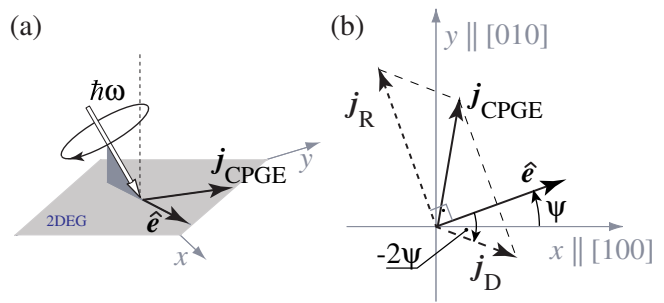}}
  \caption{ Geometry of CPGE measurements (a) and angular dependence of the CPGE current
	(b). } \label{fig3}
\end{figure}

The circular photogalvanic effect is  another  phenomenon which 
links the current to the spin splitting
in heterostructures. The CPGE is a result of selective
photoexcitation of carriers in $\bm k$-space with circularly
polarized light due to optical selection rules.\cite{review2003spin,GanichevPrettl,Ivchenkobook2,PRL01}
 In (001)-grown two
dimensional structures the CPGE current is generated by oblique
illumination of the sample with circularly polarized light only
(see Fig.~\ref{fig3}~(a)).\cite{footnote1} 
The CPGE current ${\bm j}_{\rm \,CPGE}$ is
proportional to the light's
circular polarization degree $P_{circ}$ and depends on the
direction of  light propagation, $\hat{\bm e}$:
\begin{equation}
\label{CPGE} {\bm j}_{\rm \,CPGE}  = C  \left(
{{\begin{array}{*{20}c}
 \beta' \hfill & -\alpha' \hfill \\
  \alpha'  \hfill & - \beta' \hfill \\
\end{array} }} \right) P_{circ} \hat{\bm e} \,\, ,
\end{equation}
where $C$ is the CPGE constant determined by the optical selection
rules and by the momentum relaxation time. This equation shows
that the CPGE current also consists of contributions proportional
to Rashba and Dresselhaus constants $\alpha'$ and $\beta'$ as
displayed in Fig.~\ref{fig3}~(b) by ${\bm j}_R$ and ${\bm j}_D$.
In contrast to the SGE, CPGE current is determined by
spin-splitting constants not only in the ground electronic subband
but also by those in the excited conduction and valence
size-quantized subbands. This is because the CPGE, caused by Drude
absorption, is formed due to virtual transitions via other subbands
and is hence determined by their spin-orbit splittings as
well. This fact is indicated in Eq.~(\ref{CPGE}) by using
primed Rashba and Dresselhaus constants $\alpha'$ and  $\beta'$.
Due to symmetry arguments discussed above Eq.~(\ref{CPGE}) has the
same mathematical form as Eq.~(\ref{SGE}) which describes the
spin-galvanic current. The physical content, however, is different
which reflected by the different scalar factors $A$ and $C$ as
well as by the different pseudo-vectors ${\bm S}$ and
$P_{circ}\hat{\bm e}$. Due to this similarity the evaluation of
the anisotropy of spin splitting obtained from CPGE follows
exactly the same procedures as those for the spin-galvanic effect.
The only step to be done is to replace the
in-plane average spin ${\bm S}$ by the pseudo-vector $P_{\rm
circ}\hat{\bm e}$.

Then in the geometry CPGE-I $\hat{\bm e}$ should be aligned along
the $x$ axis parallel to the $[100]$-direction ($\Psi = 0$) and the
ratio of the currents measured along the $x$ and the $y$ axes
gives
\begin{equation}
\label{RDgeomI_CPGE} \frac{\alpha'}{\beta'} =\left|
\frac{j_y(\hat{\bm e} \parallel x)}{j_x(\hat{\bm e} \parallel x)}
\right| \,\, .
\end{equation}

As for the geometry SGE-II for the geometry CPGE-II two
measurements need to be carried out, one with the $\hat{\bm e}$
along  the $y^\prime$-axis ($\Psi =  \pi/4$), and the second with
$\hat{\bm e}$ along the $x^\prime$-axis  ($\Psi= -\pi/4$). In both
cases the total current should be measured  perpendicularly to
$\hat{\bm e}$ and the R/D-ratio can be obtained from
\begin{equation}
\label{RDgeomII_CPGE}
\frac{\alpha'}{\beta'} = \frac{1+r^\prime}{1-r^\prime}\,\,\,,
\end{equation}
where
\begin{equation}
\label{RDgeomII_aCPGE} r^\prime = \left|  \frac{j_{x^\prime }(\hat{\bm
e} \parallel y^\prime)}{j_{y^\prime }(\hat{\bm e} \parallel
x^\prime)} \right| \,.
\end{equation}
Obviously, the CPGE-II method  again does not distinguish between  Rashba and
Dresselhaus terms.

\subsection{Experimental}

\begin{figure}
	\centerline{\epsfxsize 86mm \epsfbox{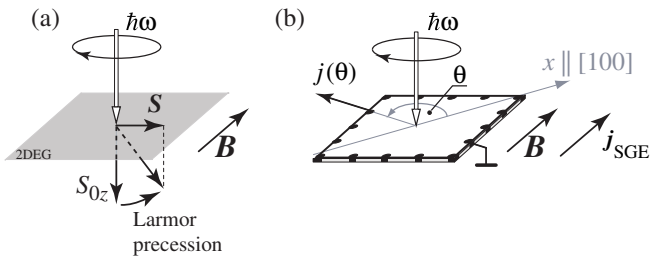}}
  \caption{ Geometry of SGE measurements under optical excitation and Hanle effect (a)
	and angular dependence of the SGE current (b).}
	\label{fig4}
\end{figure}

As shown above the experimental access to the spin-orbit coupling
parameters $\alpha$ and $\beta$ is provided by mapping the current
in the plane of the heterostructure. Basically it is sufficient to
measure  either the spin-galvanic current or the circular
photogalvanic effect in one of the geometries introduced above.
However, to have a cross check and to increase the accuracy,
redundant measurements for a number of crystallographic directions
are desirable. The  component of the total current along any
in-plane direction given by the angle $\theta$ can be written as a
sum of the projections of $\mbox{\boldmath$j$}_R$ and
$\mbox{\boldmath$j$}_D$ onto this direction, following
\begin{equation}
\label{Ch7eq3} j (\theta ) = j_D\cos (\theta +
\Psi) + j_R\sin (\theta - \Psi).
\end{equation}
Such measurements can be performed with a sample having large
number of contacts around the edges of the specimen as shown in
Fig.~\ref{fig4} (b). The current is collected from opposite contacts one-by-one
at different angles $\theta$.
%

\subsubsection{Spin-galvanic effect}

For the method based on the spin-galvanic effect one should know
the direction of the average spin in the geometry SGE-I or the
ratio between ${\bm S} \parallel x^\prime$ and ${\bm S} \parallel
y^\prime$ in the geometry SGE-II (see Eq.~(\ref{RDgeomII})). This
is, however, not an easy task.

Using optical spin orientation and an external magnetic field $\bm B$ to generate the SGE,
the necessary
information about the average spin can be obtained from the
direction of $\bm B$. The non-equilibrium in-plane
average spin {\boldmath$S$} is prepared in the following way:
Circularly polarized light at normal incidence on the QW plane
polarizes the electrons in the lowest conduction subband
resul\-ting in  monopolar spin orientation  in the $z$-direction,
$S_{0z}$ in Fig.~\ref{fig4} (a). An in-plane magnetic field
rotates the spin to the QW plane. The competition of  Larmor
precession and spin relaxation   results in the appearance of a 
steady state non-equilibrium in-plane average spin ${\bm S}
\propto {\bm \omega}_L \tau_s$, where ${\bm \omega}_L \propto {\bm
B}$ is the Larmor frequency and $\tau_s$ the spin relaxation
time.~\cite{footnote2} Finally the in-plane average spin ${\bm S}$
causes the spin-galvanic effect (Fig.~\ref{fig4} (b)). Applying a
magnetic field {\boldmath$B$} along $y$ we realize the situation
similar the method SGE-I, and orienting {\boldmath$B$} along
$x^\prime$ and $y^\prime$ we obtain the geometry SGE-II. To obtain
the R/D-ratio, however we should take into account a possible
anisotropy  of the spin relaxation.

Now let us consider, for instance, the geometry, where the 
magnetic field is oriented along $y$. The steady state spin
components can be decomposed by projections on the axes $x^\prime$
and $y^\prime$ where the spin relaxation rate tensor is diagonal
with the following components:
\begin{equation}\label{S}
S_{x'} = - {\omega_L \over \sqrt{2}} \tau_{sx'} \: S_{0z}, \qquad
S_{y'} = {\omega_L \over \sqrt{2}} \tau_{sy'} \: S_{0z} \,\, ,
\end{equation}
where $\tau_{s \, i}$ is the relaxation time for the $i$-th spin component.

Then it follows from Eqs.~(\ref{SGE}) and~(\ref{S}) that the ratio
of the currents is given by~\cite{footnote3}
\begin{equation}\label{ratio_100}
\frac{j_{y}}{j_{x}} = { \alpha (\tau_{sx'} + \tau_{sy'}) + \beta
(\tau_{sx'} - \tau_{sy'}) \over \beta (\tau_{sx'} + \tau_{sy'}) +
\alpha (\tau_{sx'} - \tau_{sy'})} \,\, .
\end{equation}

In contrast to the geometry SGE-I
the ratio of the currents $j_y/j_x$
is not simply equal to $\alpha/\beta$ but is a function
of the anisotropy of spin relaxation times.

There are two main spin relaxation mechanisms in III-V compound
based heterostructures: the precessional one proposed by
D'yakonov and Perel' and the scattering-induced mechanism of
Elliott and Yafet. While Elliott-Yafet spin relaxation is
isotropic in the heterostructure plane, the D'yakonov-Perel'
mechanism is anisotropic as shown in
Refs.~\onlinecite{Averkiev02pR271,GolubKochereshko}. This
difference affects the analysis of the SGE measurements and, on
the other hand, allows one to discriminate the dominant spin
relaxation mechanism in the studied system. In the case of
a dominant  Elliott-Yafet  mechanism spin relaxation is isotropic, therefore
$\tau_{sy'} = \tau_{sx'}$ and the
spin points always normal to the magnetic field ($\bm S \bot \bm B$).
Thus,   the ratio of the Rashba and
Dresselhaus spin-splittings in the geometry SGE-I is given by
\begin{equation}\label{ratio_EY}
{\alpha \over \beta} =  {j_{y}(\bm B \parallel y) \over j_{x}(\bm B \parallel y)} \,\, .
\end{equation}
If spin relaxation is governed by the D'yakonov-Perel' mechanism,
however, spin relaxation is proved to be anisotropic
~\cite{Averkiev1999p15582,Averkiev02pR271},
\[
{\tau_{sy'} \over \tau_{sx'}} = \left( {\alpha - \beta \over \alpha
+ \beta} \right)^2\,\,.
\]
This anisotropy needs to be taken into account yielding
\begin{equation}\label{ratio_DP}
{\alpha \over \beta} =  {(r'' +1)^{1/3} + (r''
-1)^{1/3} \over (r'' +1)^{1/3} - (r'' -1)^{1/3}} 
\end{equation}
where
\begin{equation}\label{ratio_DP1}
r'' = {j_{x}(\bm B \parallel y) \over j_{y}(\bm B \parallel y)}\,.
\end{equation}

Using the same argument we obtain $\alpha/\beta$ for $\bm B$ directed
along $x^\prime$ or $y^\prime$:
\begin{equation}\label{ratio_DP_main_axes}
{j_{x^\prime}(\bm B \parallel x^\prime) \over j_{y^\prime}(\bm B \parallel y^\prime)} = \left( {\alpha - \beta \over \alpha + \beta} \right)^3 \,.
\end{equation}


\subsubsection{Circular photogalvanic effect}

The method based on CPGE measurements needs the knowledge of the light propagation
direction which can be easily controlled.
Then the R/D-ratio is directly obtained from the current
measured in the geometry CPGE-I or CPGE-II making use of 
Eq.~(\ref{RDgeomI_CPGE}) and Eq.~(\ref{RDgeomII_CPGE}), respectively.

\section{Samples and experimental set-up}

The experiments are  carried  out on (001)-oriented $n$-type
GaAs/AlGaAs and InAs/AlGaSb heterostructures. Quantum well
structures and single heterojunctions with various widths, 
carrier densities and mobilities are grown by molecular-beam
epitaxy. Parameters of investigated samples are given in Table~I.
The sample edges are oriented along the [1$\bar{1}$0] and [110]
crystallographic axes. Eight pairs of contacts on each sample
allow us to probe the photocurrent in different directions (see
Fig.~\ref{fig4} (b)).
%

To excite spin photocurrents we use the radiation of  a high power pulsed terahertz
molecular laser optically pumped by a TEA-CO$_2$ laser.
The linearly polarized radiation at a
wavelength of 148\,$\mu$m with  10\,kW power is mo\-dified to
circular polarization by using a $\lambda/4$ quartz  plate.
The terahertz radiation induces free carrier  (Drude) absorption
in the lowest conduction subband $e1$.
The photocurrent ${\bm j}(\theta)$ is measured at room
temperature in unbiased structures via the voltage drop across a
50~$\Omega$ load resistor in a closed circuit configuration with a
fast storage oscilloscope.~\cite{review2003spin} The measured
current pulses of 100~ns duration reflect the corresponding
laser pulses.

In measurements of the spin-galvanic effect the samples are
irradiated along the growth direction and an external magnetic
field of $B=0.3$\,T is applied in the plane of the structure. In
experiments on the circular photogalvanic effect we use oblique
incidence and zero magnetic field. When investigating spin
photocurrents one should take into account that optical excitation
may generate other currents which are not the result of spin
splitting and, simultaneously, are not sensitive to switching
from right- to left-handed polarization. In the spin-galvanic
effect set-up magneto-gyrotropic effects~\cite{GanichevPrettl,BelkovJPCM,Naturephys06} may play an essential
role and in the circular photogalvanic effect set-up the linear
photogalvanic effect and the photon drag may be of importance.
These effects depend on the symmetry of the material in a
different way than spin photocurrents. Therefore in most cases it
is possible to choose a crystallographic orientation and an
experimental geometry where only spin photocurrents occur. However
for the described experiments on Rashba/Dresselhaus spin-splitting
measurements of the photocurrent in various crystallographic
directions are needed. Therefore  the total current results from
the sum of different effects. In spite of this fact spin
photocurrents can be recognized by their dependence on the
helicity of the exciting radiation. Indeed, only spin
photocurrents change their direction when the state of
polarization of radiation is switched from right- to left-handed
or vice versa. This allows one to extract the spin photocurrent
contribution from the total photocurrent. Thus, we detect the
photocurrent response  for right- ($\sigma_+$) and left-
($\sigma_-$) handed circularly polarized radiation and extract
spin photocurrents caused by spin splitting after eliminating
current contributions which remain constant for $\sigma_+$ and
$\sigma_-$ radiation. The helicity dependent photocurrent $j$ we discuss here is
obtained as $j=\left(j_{\sigma_+}-j_{\sigma_-}\right)/2$. We note
that in most cases  spin photocurrents are larger or of the same
order of magnitude as the background current
(see~\cite{review2003spin,GanichevPrettl,BelkovJPCM}) which makes  this
procedure possible.

\section{results and discussion}

To obtain the R/D-ratio the spin photocurrents are measured for
either a fixed orientation  of the in-plane magnetic field in the
case of  the spin-galvanic effect or  a fixed direction of  light
propagation in the case of the  circular photogalvanic effect.
Correspondingly $\bm B$ or $\hat{\bm e}$ are oriented along one of
the three particular crystallographic directions, namely [100],
[110] or [1$\bar{1}$0]. Then, the current  is picked-up  from the
opposite contact pairs which yields a current distribution with
respect to the crystallographic axes. The experimentally obtained
currents measured in different directions, given by the angle
$\theta$, are presented in polar coordinates.

\subsection{InAs low dimensional structures}

\begin{figure}
	\centerline{\epsfxsize 86mm \epsfbox{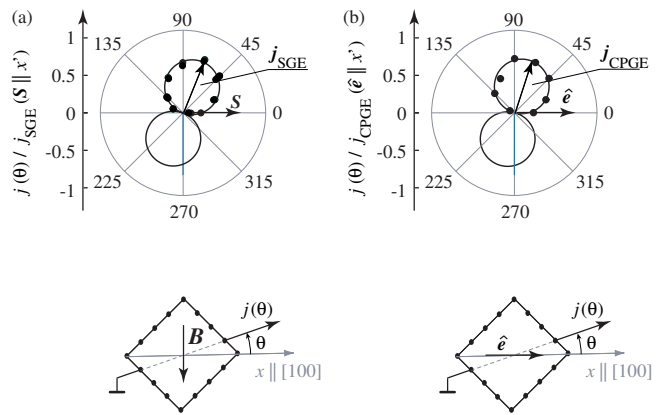}}
	\caption{ Azimuthal dependences of spin-galvanic (a) and
circular photogalvanic (b) photocurrents, j($\theta$). Data are obtained in
an   $n$-type InAs single QW (sample~1)  applying SGE-I and CPGE-I
geometries. The magnitude of the current measured at radiation
power of 10\,kW is normalized to the current maximum 
detected in $y'$-direction
($j_{\rm \,max}=j_{\rm \,SGE}\left(\bm S \parallel x^\prime \right)$)
for SGE and
($j_{\rm \,max}=j_{\rm \,CPGE}\left(\bm \hat{\bm e} \parallel x^\prime \right)$),
 respectively. Bottom panels show
geometries of corresponding experiments.} \label{fig5}
\end{figure}

Figure~\ref{fig5}~(a) shows the result for a InAs QW structure obtained
by the spin-galvanic effect for the  magnetic field oriented along
a cubic axis [010]. 
In the InAs QW structure investigated here,
the  Elliott-Yafet mechanism dominates spin
relaxation~\cite{Takeuchi1999p318} which is
isotropic.~\cite{Averkiev02pR271} Thus the anisotropy of spin
relaxation can be neglected and the in-plane average spin
{\boldmath$S$} of photoexcited carriers is perpendicular to
{\boldmath$B$}. Therefore the ratio of Rashba and Dresselhaus
currents can be  directly read off from the left hand side of
Fig.~\ref{fig5}. Helicity dependent Photocurrent flowing along the magnetic field,
$j(\pi/2)$, and therefore normal to the in-plane average spin is
$j_R$ and the current measured in the direction perpendicular to
the magnetic field $j(0)$ is $j_D$. The R/D-ratio
$\alpha/\beta = j_R/j_D = 2.1 \pm 0.3$. Moreover, using this
value for $\alpha/\beta$ all data on the
Fig.~\ref{fig5}~(a) can be fitted by Eq.~(\ref{Ch7eq3}) by one
fitting parameter corresponding to the strength of the current
being independent of the angle $\theta$. From Fig.~\ref{fig5} (a)
it is seen that the current $j(\pi/2)$ in the direction normal  to
the average spin is larger than $j(0)$. This unambiguously
shows that the Rashba contribution dominates the spin splitting.

We also obtained the R/D-ratio applying the circular photogalvanic
effect by oblique incidence of the radiation in a plane containing
the [100] axis (Fig.~\ref{fig5} (b)). Figures~\ref{fig5} (a) and
(b) demonstrate the similarity of the results obtained by
the different effects. The photogalvanic data can be fitted by
Eq.~(\ref{Ch7eq3}), too,  and yield almost the same ratio
$j_R/j_D=j(\pi/2)/j(0)=\alpha'/\beta'=2.3$.
The fact that the R/D-ratio obtained by CPGE is close to
that extracted from the SGE may be attributed either to to always
the
dominating role of the first subband in the formation of the CPGE
due to Drude absorption or to the overweighting  contribution of
the optically induced SGE.~\cite{footnote1}
We applied this method to two other InAs structures. The growth condition of these
samples were quite similar to the first one. Therefore, as
expected, the Rashba to Dresselhaus spin-splitting ratios are obtained
very close to each other (see Table~I).

\begin{figure}
	\centerline{\epsfxsize 86mm \epsfbox{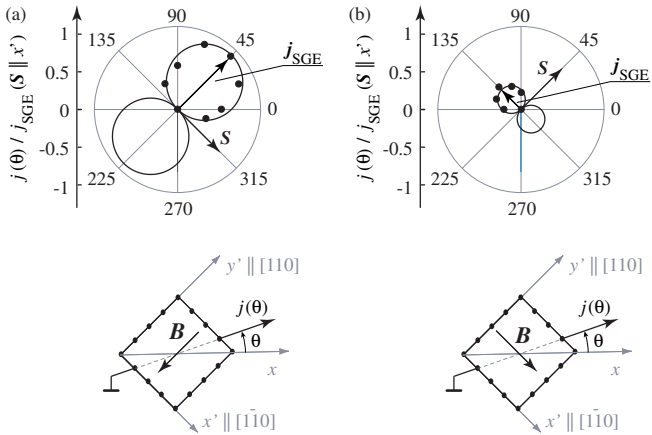}}
  \caption{Azimuthal dependences of  spin-galvanic photocurrent
	measured in $n$-type InAs single QW (sample~1) applying  SGE-II
	geometries. The magnitude of the current measured at radiation
	power of 10\,kW is normalized to the current maximum 
	($j_{\rm \,max} = j_{\rm \,SGE} \left(S\parallel x^\prime\right)=20 \mu$A) 
	detected in $y^\prime$-direction obtained in the geometry (a). Bottom panels show
	geometries of corresponding experiments.} \label{fig6}
\end{figure}

The geometry of the experiment shown so far is sufficient to
extract the R/D-ratio and to conclude on the dominating mechanism
of the spin splitting. However to demonstrate the self-consistency
of the method we performed measurements in the  geometry SGE-II as
well. Figure~\ref{fig6}    presents data obtained by the
spin-galvanic effect. 
Applying the magnetic field   along the
[110] or [1$\bar{1}$0] directions we obtain current distributions
plotted in Fig.~\ref{fig6} (a) and (b). It is seen that, as
expected from Eq.~(\ref{j_total}), for these geometries the
current always reaches its maximum perpendicularly to the
in-plane average spin. The magnitudes of the helicity dependent current in
Figs.~\ref{fig6}~(a) and (b)  are substantially different. Note
that the data in Figs.~\ref{fig6}~(a) and  (b) is normalized to
$j_{SGE}(\bm S \parallel x^\prime) = j(\pi/4)$ measured in the
geometry of Fig.~\ref{fig6} (a). This is the maximum possible
current because Rashba and Dresselhaus contributions point  in the
same direction in this geometry (Fig.~\ref{fig2}~(d)). The ratio
of Rashba and Dresselhaus currents can be evaluated from the
currents   $j(\pi/4)$ in Fig.~\ref{fig6} (a)  ($j \propto (\alpha
+ \beta)$) and  $j(3\pi/4)$ in Fig.~\ref{fig6} (b) ($j \propto
(\alpha - \beta)$). Knowing that the Rashba spin-splitting is
dominant, this procedure gives the same result as obtained by the
first geometry: $j_R/j_D=2.1$ demonstrating the self-consistency
of the method described. Investigation of the CPGE response in the
similar geometry ($\hat{\bm e} \parallel \langle 110 \rangle$)
gives the same result. As  addressed above the dominance  of
the Rashba term cannot be deduced solely from these measurements.

The R/D-ratio  of 1.6 -- 2.3 agrees well with theoretical
results~\cite{Lommer88p728} which predict a dominating Rashba
spin-orbit coupling for InAs QWs and is also consistent with
 expe\-riments.\cite{Knap1996p3912,Nitta97p1335} For InGaAs
QWs, having similar sample parameters as the structures
investigated here, $\alpha/\beta$ ratios were obtained from weak
anti-localization experiments~\cite{Knap1996p3912} and $\bm k
\cdot \bm p$ calculations.\cite{Pfeffer99pr5312} The corresponding
values ranged between 1.5~-~1.7 and 1.85, respectively. These
results are in good agreement with our findings. The R/D-ratio has
previously been estimated from fits of magnetotransport and Hanle
experiments.~\cite{Knap1996p3912,Miller03p076807}
In contrast
to these works our method allows to measure directly the
relative strength of Rashba and Dresselhaus terms and does not
require any additional theoretical estimations.

\subsection{GaAs low dimensional structures}

The R/D-ratio for GaAs quantum wells is obtained applying both
SGE-I and CPGE-I methods. The results are presented in
Table~I. Measurements show that the Rashba constant $\alpha$ is
larger than the Dresselhaus constant $\beta$ in all studied
samples.

Let us first discuss the results for the single heterojunction. In
CPGE measurements in geometry I one obtains $j(\pi/2)/j(0) = 7.6$
which yields consequently $\alpha / \beta = 7.6$. The fact that in
single heterojunctions the Rashba contribution is much larger  is
not surprising because, on the one hand, such structures, due to
the triangular confinement potential, are strongly asymmetric and,
on the other hand, the $\bm k$-linear Dresselhaus terms in wide
structures like here are reduced. Measurements of the
spin-galvanic effect in such structures give the ratio of the
currents measured for a magnetic field along the $y$-axis:
$j(\pi/2)/j(0) = 2.65$. It is well known that the dominant spin
relaxation mechanism in GaAs is the D'ykonov-Perel' relaxation,
therefore the R/D-ratio should be calculated  applying the
relevant Eqs.~(\ref{ratio_DP}) and~(\ref{ratio_DP1}). We then obtain
$\alpha/\beta = 7.6$ which is in agreement with the value obtained
from CPGE measurements and demonstrates applicability of the SGE
method also for a dominating D'yakonov-Perel' mechanism (see
Table~I). 

Now we compare spin splittings in the single heterojunction and
the 8.2~nm wide QW. It is seen that the absolute value of the
R/D-ratio is larger for the heterojunction than for the
 QW, and the signs of the ratios are opposite. The change of
the absolute value can be attributed to two facts. On the one
hand, increase of the electron confinement increases the
Dresselhaus constant $\beta$. On the other hand, SIA in the
heterojunction is higher which means larger Rashba constant
$\alpha$. Self-consistent solution of Schr{\"o}dinger and Poisson
equations for electron wave functions in our  heterojunction  and
in the 8.2~nm wide QW shows that the wave function is indeed stronger localized in
the QW than in the heterojunction. The relative sign of
R/D-ratios is obtained from directions of the currents ${\bm j}_R$
and ${\bm j}_D$. Measurements show that the directions of the
Dresselhaus current ${\bm j}_D$ in both samples remains the same,
while the direction of the Rashba current ${\bm j}_R$ is reversed in
the 8.2~nm wide QW. Conservation of
the ${\bm j}_D$ sign is obvious because the sign of the
Dresselhaus spin-splitting is determined by the bulk properties of
the material and, hence, is the same in both GaAs based systems.
The reversing of ${\bm j}_R$, caused by SIA, follows from the fact
that built-in electric fields, determining SIA, have opposite
directions in these structures resulting in opposite signs of the
Rashba constants $\alpha$.

In the 15~nm and 30~nm wide GaAs QWs we controllably varied the
strength of SIA. This has been achieved by doping at different
distances from the QW. For quantum wells of the same width,
asymmetrical and symmetrical modulation doping yields  larger and
smaller strengths of SIA, respectively. At the same time for a
given QW width, the strength of the Dresselhaus spin-splitting is
constant because it is independent of the doping position. Thus,
variation of the doping distance should affect the R/D-ratio.
Indeed, for more symmetric QW structures of both 15~nm and 30~nm
width the ratio is smaller (see Table~I). Opposite signs of the
R/D-ratio for these samples as well as for previously discussed
GaAs samples, are consistent with the direction of built-in
electric fields obtained from solving Schr{\"o}dinger and Poisson
equations.

We emphasize that the observed 
sign reversal of the R/D-ratio upon changing of the $\delta$-doping  plane  in the
heterostructure  opens a way of QW grows with controllable spin
splitting.
%
Our results demonstrate that SIA can be
tuned in a wide range with $\alpha$ both positive and negative.
Measuring the Rashba photocurrents in the geometries SGE-I or
CPGE-I as a function of doping position in structures of 
same width allows one to find the sample with zero Rashba photocurrent,
corresponding to $\alpha = 0$. Applying geometry SGE-II or
CPGE-II for the sample one can find zero photocurrent which,
according to Eqs.~(\ref{RDgeomII}) and (\ref{RDgeomII_CPGE}), means
that the condition of equal Rashba and Dresselhaus
constants~\cite{Schliemann03p146801} ($\alpha = \beta$),
important for spintronics applications, is realized.

\section{Summary}

To summarize, we investigate Rashba and Dresselhaus spin-orbit
splittings in various III-V material based 2D structures at room
temperature applying spin photocurrents. We use the SGE and the CPGE and
demonstrate self-consistency of the obtained results. In all
investigated samples the Rashba contribution dominates over the
Dresselhaus one. We emphasize that our measurements give the
necessary feedback 
for structures with
equal Rashba and Dresselhaus spin-splittings or perfectly
symmetric structures with zero Rashba constant.

\section*{Acknowledgements}
We thank E.L.~Ivchenko and S.A.~Tarasenko, as M.~Koch and T.~Kleine-Ostmann
for helpful discussions. The high quality InAs quantum wells
were kindly provided by J.~De~Boeck and G.~Borghs from IMEC
Belgium. This work is supported by the Deutsche Forschungsgemeinschaft via
Collaborative Research Center SFB689 and GRK638, 
Russian President grant for young scientists, RFBR, and Russian Science
Support Foundation, RAS and HBS.

\begin{widetext}

\begin{table}
\caption{Parameters of samples and  measured R/D ratios.  Mobility
and electron sheet density data are obtained at 4.2~K in the dark.
We note that  R/D-ratios are extracted from SGE measurements on a
limited number of samples because in some structures the
Dresselhaus SGE-current has been masked by magneto-gyrotropic
currents.~\protect \cite{BelkovJPCM}}

\begin{tabular}{ccccccccc}

\hline
\,\,\,\, sample &  \,\,\,\, material & \,\,\,\, QW width & \,\,\,\, spacer 1 & \,\,\,\, spacer 2 & \,\, \,\,mobility  &  \,\,\,\, density &  \,\,\,\, $\alpha/\beta$ &  \,\,\,\, $\alpha'/\beta'$\\
        &          &     \AA\     &   \AA\        &   \AA\        & \,\,\,\,cm$^2$/Vs  & \,\,\,cm$^{-2}$ & SGE & CPGE \\
\hline
\#1  & InAs/AlGaSb     & 150      &   -   &  -          & $3.0 \cdot 10^5$   & $8 \cdot 10^{11}$ & 2.1 & 2.3\\
\#2  & InAs/AlGaSb     & 150      &  -    &  -          & $2.0 \cdot 10^5$   & $1.4 \cdot 10^{12}$ & --  & 1.8\\
\#3  & InAs/InAlAs     & 60      &  -    &  75        & $1.1 \cdot 10^5$   & $7.7 \cdot 10^{11}$ & --  & 1.6\\
\#4  & GaAs/AlGaAs     & $\infty$ &  700 &   -         & $3.5 \cdot 10^6$   & $1.1 \cdot 10^{11}$ & 7.6 & 7.6 \\
\#5  & GaAs/AlGaAs     & 82       &  50 &  50        & $2.6 \cdot 10^6$   & $9.3 \cdot 10^{11}$ & -4.5 & -4.2\\
\#6  & GaAs/AlGaAs     & 150      &  600 & 300        & $1.0 \cdot 10^5$   & $6.6 \cdot 10^{11}$ & --  & -3.8\\
\#7  & GaAs/AlGaAs     & 150      &  400 & 500        & $2.6 \cdot 10^5$   & $5.3 \cdot 10^{11}$ & --  & -2.4\\
\#8  & GaAs/AlGaAs     & 300      &  700 &   -         & $3.2 \cdot 10^6$   & $1.3 \cdot 10^{11}$ & --  & 2.8\\
\#9  & GaAs/AlGaAs     & 300      &  700 & 1000       & $3.4 \cdot 10^6$   & $1.8 \cdot 10^{11}$ & --  & 1.5\\
\hline
\end{tabular}

\end{table}

\end{widetext}


\begin{thebibliography}{99}
\bibitem{footnote2}The measurements should be carried out at low
magnetic field strengths so that $\omega_L \tau_s < 1$ and, thus,
the Hanle denominator in ${\bm S}$ can be neglected.
\bibitem{footnote3} Here we neglect the
anisotropy of the $g$-factor in the plane of the structure
assuming it to be small, as it has been demonstrated in an
asymmetric GaAs QWs of 10~nm width in Ref.~\protect \cite{g_aniz} where
the anisotropy was about ten percents.
\end{thebibliography}
\end{document}